\documentclass[prc,aps,floatfix,nofootinbib,preprint]{revtex4} 
\usepackage{graphicx} 
\usepackage{amssymb} 
\usepackage{amsmath} 
\usepackage{amsfonts} 
\usepackage{color}

\usepackage{ulem}

\def\be{\begin{equation}} 
\def\ee{\end{equation}}

\renewcommand{\vec}[1]{\mbox{\boldmath $#1$}}

\begin{document}

\title{Diabatic Hamiltonian matrix elements made simple}

\author{K. Hagino}
\affiliation{ 
Department of Physics, Kyoto University, Kyoto 606-8502,  Japan} 

\author{G.F. Bertsch}
\affiliation{ 
Department of Physics and Institute of Nuclear Theory, Box 351560, 
University of Washington, Seattle, Washington 98915, USA}

\begin{abstract}
With a view to applying the Generator Coordinate Method  to large
configuration spaces, we propose a simple approximate formula to compute
diabatic many-body matrix elements without having to evaluate  
two-body interaction matrix elements. The method is illustrated
with two analytically solvable Hamiltonians based on the
harmonic oscillator.
\end{abstract}  

\maketitle

\section{Introduction}

In nuclear physics we  use transparent models not only to understand the 
observed characteristics of nuclear spectra and dynamics but also to 
provide checks and guidance on the complex computer codes that produce
the quantitative theory of their properties.  
The Lipkin model is a typical example \cite{RS}.

In this spirit we derive
and apply a simple formula, Eq. (\ref{orbital-eq1}) below, for calculating matrix elements between
configurations in the generator 
coordinate method (GCM).  
In GCM, many Slater determinants are linearly superposed 
to describe low-energy dynamics of many-body systems \cite{RS,RG87,RRR18}. 
Since those Slater determinants are not orthogonal, one needs to calculate 
their overlaps and Hamiltonian matrix elements.
To this end, the Hamiltonian matrix elements are computed using the
generalized Wick's theorem (GWT)\cite{BB} as was first done in Ref. \cite{bo90}.

However, the GWT requires
a full knowledge of the Hamiltonian in a Fock-space representation.
In contrast, Eq. (\ref{orbital-eq1}) below 
only needs the overlap matrix elements of the orbitals and the
orbital energies calculated from the single-particle Hamiltonian.
It is an approximate formula that can be used with any of the theories in which an effective
single-particle Hamiltonian can be defined.  This includes theories
based on energy functionals such as Skyrme, Gogny and the BCPM\cite{ba13}, as well as  
those making use of the Strutinsky treatment of the total energy.   
In this paper, we shall apply the formula to  
two simple models, namely, the translational motion in a harmonic oscillator 
potential and 
a simplified Nilsson model. 

\section{Orbital energy formula}

We first summarize the usual computational procedure.  The
starting point is a configuration space of many-particle wave
functions that have the form of Slater determinants of orbitals
in some common basis.\footnote{The extension to include pairing
by the Hartree-Fock-Bogoliubov equation is beyond the scope of
this paper.}  The 
Hartree-Fock method reduces a Hamiltonian with two-particle interactions
to a one-particle Hamiltonian with the orbitals as its eigenstates.\footnote{
The method can easily be extended to Hamiltonians with three-particle 
and higher terms by the same reduction of the higher-order density
matrices\cite{ro12} to what has been 
done in the Hartree-Fock approximation.} The orbitals are obtained by
minimizing the single-particle Hamiltonian or energy functional.
The space of configurations in the GCM is expanded by introducing a
collective coordinate.  This is often implemented by adding a term
in the Hamiltonian to constrain the expectation value of some
single-particle operator, but the coordinate can also be defined
in other ways.  In the examples below, it is convenient to introduce
the coordinate directly into the orbital wave functions.

Assuming that the constraint is described by a single parameter $q$,
the task is to calculate the Hamiltonian matrix element between a
configuration
$|\Psi_q\rangle$ and some other configuration $|\Psi_{q'}\rangle$.  The GWT method 
can be derived from  Thouless's theorem
\cite{thouless1960} which is used to express the second wave function 
the orbital basis of the first one.  That is,
\be
|\Psi_{q'}\rangle = {\cal N} \prod_{h} \left(1 + \sum_pc_{ph}a^\dagger_p
a_h\right) |\Psi_q\rangle.
\ee
Here $h$ and $p$ are occupied and unoccupied orbitals in the basis of  
$|\Psi_q\rangle = \prod_h a^\dagger_h |\rangle$.
The Fock-space operators are expressed
in the $q$ orbital basis  and $\cal N$ is a normalization 
constant given by
\be
{\cal N} = (\det M)^{-1/2}
\ee
where $M_{h,h'} = \delta_{h,h'} + \sum_p
c_{ph} c_{ph'}$.
With all of the operators expressed in
the same basis it is easy to apply the ordinary Wick's theorem to calculate
arbitrary many-body matrix elements.  

To derive the orbital energy formula we place the reference configuration
midway between the two states $|\Psi_q\rangle$ and $|\Psi_{q'}\rangle$.  Changing the notation
somewhat, the many-particle configurations can be written 
\be
|\Psi_{\pm q} \rangle = {\cal N^{\pm}} \prod_{h} 
\left( 1 + \sum_p  c_{ph}^{\pm} a^\dagger_p a_h\right)
|\Psi_0\rangle,
\label{wf-z}
\ee
where  $|\Psi_0\rangle$ is the reference configuration and 
$|\Psi_{\pm q}\rangle$ are the 
end configurations.   The overlap between the two end configurations is given by
\be
\langle \Psi_{-q} | \Psi_{+q}\rangle 
= {\cal N}^{-}{\cal N}^{+} \det \left|\delta_{h,h'}
+ \sum_p c^{+}_{ph} c^{-}_{ph'}\right|.
\ee
The object is to calculate the matrix element of the 
Hamiltonian between them.  

We write the effective
Hamiltonian in Fock space as
\begin{eqnarray}
\hat H &=& E_0 + \sum_{p}\varepsilon_p a^\dagger_p a_p - \sum_{h}\varepsilon_h
a_h a^\dagger_h + \nonumber \\
&&+ \sum_{i\ne j} H_{ij}a^\dagger_i a_j +\sum_{ij,k\ell} v_{ij,k\ell} a^\dagger_i a^\dagger_j 
a_\ell 
a_k.
\label{Hfock}
\end{eqnarray}
Here $E_0 = \langle \Psi_0 |\hat H | \Psi_0 \rangle $ 
is the expectation value of the
Hamiltonian in the reference configuration.  
The effective single-particle
Hamilton $H_{ij}$ has been split into diagonal contributions  $H_{ii} =
\varepsilon_i$ and off-diagonal ones appearing on the second line.  
The last term is the effective two-particle interaction.  It contains all
the interaction matrix elements not present in the other terms.
We will argue that its contribution
is small when  the matrix elements are calculated between states related
by a diabatic transformation.  

In a diabatic evolution, each orbit
is transformed independently according to a local single-particle
operator such as the quadrupole operator $\hat Q = z^2 - (x^2+y^2)/2$.
Under those conditions the transition density $\langle \Psi_q| \hat \rho |
\Psi_{q'}\rangle =
\delta \rho$ is collective with a strong local component proportional to
$d \rho(\vec r)/dq$ in the coordinate
space representation.  The mean field $V_0$ behaves similarly.
The interaction can then be parameterized
in the separable form
\be
v_{ij,k\ell} = \frac{\kappa}{2} \sum_{{\lambda}\mu} O^{{\lambda}\mu *}_{ik} 
O^{{\lambda}\mu}_{j\ell}
\label{vsep-1}
\ee
with single-particle matrix elements 
\be
O^{{\lambda}\mu}_{ij} = \int d^3 r \,  \phi^*_i(\vec r) \phi_j(\vec r) \frac{d V_0}{d r}
Y_{{\lambda}\mu}(\hat{\vec{r}}).
\label{vsep-3} 
\ee
The strength $\kappa$ is determined by the self-consistency
condition \cite{bohr-mottelson,bertsch1983}
\be
\kappa^{-1}=\int_0^\infty
r^2dr\,\frac{d\rho_0(r)}{dr}\frac{dV_0(r)}{dr}. 
\label{self-consistent}
\end{equation}
In calculating the interaction between many-particle configurations, one 
includes only the 
direct term in the interaction; exchange effects are implicit in the
single-particle potential $V_0$.  Except for $E_0$, the resulting Hamiltonian
can be
generated entirely in terms of the effective single-particle mean field and
its variation under the diabatic transformation.
Interactions of this form  have been applied
not only to  collective excitations \cite{su81,tagami2013} but even to treat
nuclear spectra as a whole \cite{alhassid}. 

We now estimate many-body Hamiltonian matrix elements using Eq.
(\ref{Hfock}-\ref{self-consistent}).  Several assumptions in 
the derivation are justified by considering the limit of a large
number of particles $N_h$ together with small displacements in the
collective coordinate.  This leads to a simplification of the L\"owdin formula 
{\cite{lowdin1955,brink1965}} 
for matrix elements requiring the minors of the normalization determinant
$\det(M)$.  The off-diagonal
minors are much smaller than the diagonal ones when the
$c_{ph}$ amplitudes are small.  Also, the diagonal ones are approximately
equal to $\det(M)$.  The Hamiltonian matrix element can then be evaluated
as
\begin{eqnarray}
\frac{\langle \Psi_{-q} | \hat H | \Psi_{+q} \rangle}{\langle \Psi_{-q}|\Psi_{+q}\rangle}
&\approx & 
E_0 + \sum_{ph} c^+_{ph} c^-_{ph} (\varepsilon_p - \varepsilon_h)
\label{full-eq}
\\
&&+ \sum_{ph} H_{ph} (c^+_{ph} + c^-_{ph})   \nonumber \\
&&+ \frac{\kappa}{2}  \sum_{ph,p'h'}\left(c^+_{ph}c^-_{p'h'}+c^-_{ph}c^+_{p'h'}
+c^+_{ph}c^+_{p'h'} +c^-_{ph}c^-_{ph}\right) 
\sum_{{\lambda}\mu}O^{{\lambda}\mu *}_{ph}O^{{\lambda}\mu}_{p'h'} \nonumber.
\end{eqnarray}
Eq. (\ref{full-eq}) can be further simplified in several ways.  First, we limit the
scope to axially symmetric diabatic transformations.  Then the last sum
can be restricted to $\mu=0$ operators, and the amplitudes can be treated as 
real quantities.  Next, we can assume
that $c^- =  -c^+$ if $q$ is not too large.  Then the last term vanishes.
Note that the cancellation requires that the separable terms in the 
interaction are even under the time-reversal transformation. 
Finally,
if the state $|0\rangle$ is a self-consistent solution of the mean-field
equation, the off-diagonal elements $H_{ph}$ vanish as well.  

Changing the notation back to the original, the resulting formula reads
\be
\frac{\langle \Psi_{q_1} | \hat H | \Psi_{q_2} \rangle}{\langle \Psi_{q_1}|
\Psi_{q_2}\rangle}
\approx E_0 - 
\sum_{ph} c^2_{ph} (\varepsilon_p - \varepsilon_h)
\label{orbital-eq1}
\ee
where the particle-hole amplitudes have been determined with respect to 
a reference configuration at $q = (q_1 + q_2)/2$.
It is remarkable that all dependence on the residual interaction has disappeared.

The formula can be easily applied to calculate the inertial parameter
associated with the collective coordinate $q$.  For small $q$ 
the amplitudes may be expanded as a power series with 
the leading term $ c_{ph} = \frac{1}{2}(q_1-q_2)\,\,  d c_{ph} / d q$ to give
\be
\frac{\langle \Psi_{q_1} | \hat H | \Psi_{q_2} \rangle}{\langle \Psi_{q_1}|\Psi_{q_2}\rangle}
\approx
E_0 - \frac{1}{4}\sum_{ph} \left(\frac{d c_{ph}}{d q}\right)^2 
(q_1-q_2)^2 (\varepsilon_p - \varepsilon_h). 
\label{orbital-eq2}
\ee
Finally, one can calculate the collective inertial parameter
$B$ by applying the formula to the plane-wave state 
\begin{equation}
{|k\rangle = \int d q |\Psi_q\rangle
e^{ikq}.} 
\label{fourier}
\end{equation}
Writing the derivative as $d c_{ph}/ d q = c'_{ph}$, 
one has
\begin{equation}
{\langle\Psi_q|\Psi_{q'}\rangle 
\sim e^{-\sum_{ph}(c'_{ph})^2(q-q')^2/2}}
\end{equation}
for small $(q-q')$.
One {thus} obtains
\be
{\frac{\langle k | \hat H | \Psi_q \rangle}{\langle k| \Psi_q\rangle}
\approx  E_0} - \frac{\sum_{ph} (c'_{ph})^2 (\varepsilon_p -\varepsilon_h)}{4\sum_{ph} (c'_{ph})^2} + 
\frac{{\sum_{ph} (c'_{ph}})^2 (\varepsilon_p -\varepsilon_h)} {4(\sum_{ph} (c'_{ph})^2)^2}
k^2.
\ee
The third term is the collective kinetic energy $k^2/2B$
with $B$ given by
\be
B = \frac {2(\sum_{ph} (c'_{ph})^2)^2}{{\sum_{ph} 
(c'_{ph}})^2 (\varepsilon_p -\varepsilon_h)}.
\ee
{This agrees with the inertia derived from GCM/GOA using sum rules
of a constraining field \cite[Eq.(22)]{fiolhais83}. }

\section{Model 1: Translational motion}

Here we consider a Hamiltonian of $N$ distinguishable particles interacting through
a translationally invariant quadratic potential {in one dimension},
\be
H =  T + v_2 \\
=\sum_{i=1}^N \frac{p_i^2}{2 m}  + \frac{m \omega_v^2}{2}\sum_{i<j} (x_i -
x_j)^2. 
\ee
The GCM algebra below generalizes the discussion of the harmonic oscillator
model in Ref. \cite[Sec. 10.5]{RS}.

The first task is to find a mean-field ground-state wave function $\Psi_0$.  
Since the particles are distinguishable, all the orbitals $\phi_i$ are the
same and the many-body ground-state configuration has the form
\be
\Psi_0 = \prod_{i=1}^N \phi(x_i).
\ee
We may assume that the orbitals are Gaussian functions centered at
$x=0$,
\be
\phi(x) = \left(\frac{\alpha}{\pi}\right)^{1/4} e^{-\alpha x^2/2},
\ee
with a common size parameter $\alpha$.  It can be 
verified later that the 
Gaussian form allows a self-consistent solution of the mean-field equations.  
The expectation value of the Hamiltonian is
\be
\langle \Psi_0 | H | \Psi_0 \rangle = N \frac{\alpha{\hbar^2}}{4 m} + N(N-1) \frac{m
\omega_v^2}{{4} \alpha}.
\ee   
The first term is from the kinetic energy operator $T$ and the second
is from the interaction $v_2$.
Minimizing this expression with respect to $\alpha$ yields
\be
\alpha = (N-1)^{1/2}\omega_v  m{/\hbar}
\ee
giving a total energy
\be
E_0 = \langle \Psi_0 | H | \Psi_0 \rangle =
N\frac{(N-1)^{1/2}}{2}{\hbar}\omega_v.
\ee

The next GCM task is to define a generator coordinate; we take it
to be a displacement of the ground-state configuration by an amount  $z$,
\be
\Psi_{z} 
=\left(\frac{\alpha}{\pi}\right)^{N/4}\prod_{i=1}^N  e^{-\alpha (x_i - z)^2/2}
{\equiv
\prod_{i=1}^N \phi_z(x_i).
}
\ee

Applications of the GCM require the overlap matrix elements  $\langle \Psi_z | \Psi_{z'}\rangle$ and the
Hamiltonian matrix elements $\langle \Psi_z | H | \Psi_{z'}\rangle$.  The overlaps
are given by 
\be
\langle \Psi_z | \Psi_{z'}  \rangle = \exp\left(-N
\frac{\alpha}{4} (z -z')^2\right).
\label{olp-m1}
\ee
Decomposing the matrix elements of the Hamiltonian into the kinetic term and
the interaction term, we have
\be
{\langle \Psi_z |T| \Psi_{z'}  \rangle= }
\left(N \frac{\alpha{\hbar^2}}{4 m} - N\frac{\alpha^2{\hbar^2}}{8 m } (z - z')^2\right)
{\langle \Psi_{z} | \Psi_{z'}\rangle}
\ee
for the kinetic term and
\be
\langle \Psi_{z} |v_2| \Psi_{z'}  \rangle=
N(N-1) \frac{m \omega_v^2}{4 \alpha} \langle \Psi_z | \Psi_{z'}\rangle,
\ee
for the interaction term.  Note that it is the same as the expectation value
in the ground state except for the overlap factor. Combining the two terms, the
Hamiltonian matrix element can be expressed 
\be
\langle \Psi_{z} | H |\Psi_{z'} \rangle = 
\langle \Psi_{z}| \Psi_{z'} \rangle \left( 
E_0 - h_2 (z- z')^2 \right)
\label{h-m1}
\ee
where
\be
h_2 = 	N\frac{\alpha^2{\hbar^2}}{8m}.
\label{h2-m2}
\ee		
{This is identical to \cite[Eq. 10.51]{RS} except for the $N$ dependence.}

 One can derive the effective mass associated with the collective
coordinate from Eq. (\ref{olp-m1}) and (\ref{h-m1}).  
It will be shown below that it is just the value
for translational motion, namely $N m$.  

Before that, we verify the orbital energy formula, Eq. (\ref{orbital-eq1}), choosing 
end configurations at $\pm z$ and the reference configuration
at $z=0$.  The $\Psi_{z}$  particle-hole amplitudes in the basis of $\Psi_0$
are calculated by taking orbital overlaps. We only need orbitals up to the second
excited state for our purposes.  The amplitudes are  
\be
{\phi_z(x) =  \psi_0(x-z)
=e^{-\alpha z^2/4}\left(\phi_0(x)   + z\left(\frac{\alpha}{2}\right)^{1/2}\phi_1(x)  + 
z^2\frac{\alpha}{ 8^{1/2}}\phi_2(x) + ...\right).}
\ee
Next we need  the orbital excitation energies, obtained from the
spectrum as
$ \varepsilon_n  = n \alpha {\hbar^2}/m$.
Then the orbital excitation formula gives 
\be
{\sum c^+_{10} c^-_{10} (\varepsilon_1 - \varepsilon_0) = N \frac{\alpha}{2}
\frac{\alpha}{m}\hbar^2\cdot\frac{1}{4}(z-z')^2},
\ee
in agreement with {Eq. (\ref{h-m1}) with Eq. (\ref{h2-m2})}.

{
Let us now discuss the associated mass term in the collective
Hamiltonian.}
We can construct a plane-wave state from the
GCM configurations {according to Eq. (\ref{fourier}). }
The overlap with $\Psi_0$ is
\be
{\langle \Psi_0 | k\rangle} 
= \left(\frac{4 \pi}{\alpha N}\right)^{1/2}
e^{-k^2/\alpha N}
\ee
and the Hamiltonian matrix element reduces to
\be
{
{\langle \Psi_0 | H|k\rangle} =
\langle \Psi_0 | k\rangle} \left(\frac{k^2}{2N m} - \frac{\alpha}{4
m}\right).
\ee
The first term in parenthesis $k^2/2N m$ is just what we expected for
a free particle of mass $m$.  The other term is the zero point kinetic energy
associated with the center-of-mass coordinate in the original GCM
configuration composed of Gaussian orbitals.

It is not generally recognized that the GCM/GOA methodology can deliver 
exact composite-particle masses for translational motion.  It was noticed
as an empirical finding in Ref. \cite{be19} that the translational masses of
nuclei calculated with a Gogny functional were close to $A m$, the number
of nucleons $A$ times the nucleon mass $m$.  

\section{Model 2: Dynamics in a deformation coordinate}

We next consider a simplified Nilsson  model to illustrate the use of the self-consistent
separable interaction Eq. (\ref{vsep-1}) to generate a two-particle interaction from
a diabatic treatment of the mean field.  
The single-particle potential is given by a deformed harmonic oscillator
potential,
\begin{equation}
V(\vec{r})=\frac{1}{2}m\omega_z^2z^2+\frac{1}{2}m\omega_\perp^2(x^2+y^2),
\end{equation}
where $\omega_z^2=\omega_0^2(1-4\delta/3)$ and
$\omega_\perp^2=\omega_0^2(1+2\delta/3)$, following the notation of Ref.
\cite[Eq 5-5]{bohr-mottelson}. 
This potential can also be expressed as
\begin{equation}
V(\vec{r})=\frac{1}{2}m\omega_0^2r^2
-\frac{{2}}{3}\delta \,m\omega_0^2 \left(z^2- (x^2 + y^2)/2\right).
\label{eq:defho}
\end{equation}
We take the parameter $\delta$ as the generator coordinate\footnote{Another parameter in
common use is $\beta_2\approx\frac{2}{3}\sqrt{\frac{4\pi}{5}}\,\delta$.}.  
 
The eigenstates of the Hamiltonian are specified by a set of quantum numbers 
$(n_x,n_y,n_z)$ together with the $z$-components of spin {and isospin}; 
their spatial
distributions are determined by $\omega_0$ and $\delta$,  combined to
give oscillator length parameters  $b_i = \sqrt{\hbar / m \omega_i}$ for
$i = {z,\perp,0}$.

For simplicity, we consider a harmonic oscillator model of the nucleus 
$^{16}$O.  Its 
8 neutrons {and 8 protons 
occupy the ``1s'' state 
with $N_{osc}=n_x+n_y+n_z=0$ and the 
three  ``1p'' states with $N_{osc}=1$.   
The density distribution is given by 
\begin{eqnarray}
\rho(\vec{r})&=&{4}\sum_k|\phi_k(\vec{r})|^2 \\
&=&\frac{{4}}{\pi b_\perp^2\sqrt{\pi b_z^2}}\left[1+2\left(\frac{x^2+y^2}{b_\perp^2}
+\frac{z^2}{b_z^2}\right)\right] \nonumber \\ 
&&\times e^{-z^2/b_z^2}e^{-(x^2+y^2)/b_\perp^2},
\label{eq:density}
\end{eqnarray}
where 
$k\equiv (n_x,n_y,n_z)=(0,0,0),~(0,0,1),~(0,1,0)$, and (1,0,0). 

To the linear order of $\delta$, the density distribution (\ref{eq:density}) 
is expanded to 
\begin{equation}
\rho(\vec{r})=\rho_0(r)+\rho_2(r)P_{2}({\cos\theta})
\end{equation}
with 
\begin{equation}
\rho_0(r)
=\frac{{4}}{(\pi b_0^2)^{3/2}}
\,\left(1+\frac{2r^2}{b_0^2}\right)e^{-r^2/b_0^2}
\end{equation}
and 
\begin{equation}
\rho_2(r)
={\frac{\delta}{3(\pi b_0^2)^{3/2}}}
\left(\frac{4r^4}{b_0^4}-\frac{2r^2}{b_0^2}\right)
\, e^{-r^2/b_0^2}. 
\end{equation}

Given the form of the potential (\ref{eq:defho}), 
the many-body Hamiltonian with a separable 
interaction reads 
\begin{equation}
H=\sum_i\left(\frac{\vec{p}_i^2}{2m}+\frac{1}{2}m\omega_0^2r_i^2\right) 
+\frac{\kappa}{2}\left(\sum_ir_i^2 P_{2}({\cos\theta_i})\right)^2. 
\label{eq:Hmb-defho}
\end{equation}
The strength $\kappa$ evaluated with Eq. (\ref{self-consistent}) is
\begin{equation}
\kappa =-{\frac{2}{3}}\,\delta m\omega_0^2
\left({\frac{4\pi}{5}}\int^\infty_0r^2dr\,r^2\rho_2(r)\right)^{-1}
=-\frac{m^2\omega_0^3}{{18}}.
\end{equation}
\begin{table}[htb] 
\begin{center} 
\begin{tabular}{c|ccc} 
\hline
\hline
$n$ & $\langle n b | n b'\rangle$  & $\langle n b |x^2 | n b'\rangle$ & $\langle n b
|p^2| nb'\rangle$ \\  
\hline 
0 & $\sqrt{\frac{2bb'}{b^2+b'^2}}$ & 
$\frac{bb'}{2}\,\left(\frac{2bb'}{b^2+b'^2}\right)^{3/2}$ &  
$\frac{\hbar^2}{b^2+b'^2}\,\left(\frac{2bb'}{b^2+b'^2}\right)^{1/2}$ \\
1 & $\left(\frac{2bb'}{b^2+b'^2}\right)^{3/2}$ &
$\frac{3bb'}{2}\,\left(\frac{2bb'}{b^2+b'^2}\right)^{5/2}$ &  
$\frac{3\hbar^2}{b^2+b'^2}\,
\left(\frac{2bb'}{b^2+b'^2}\right)^{3/2}$ \\
\hline
\hline
\end{tabular} 
\caption{
Overlaps $\langle n b | n b'\rangle {\equiv} \langle\phi_n(b)|\phi_n(b')\rangle$ and
needed   
matrix elements for one-dimensional harmonic oscillator wave functions with oscillator 
lengths $b$ and $b'$. The oscillator quanta are denoted by $n$.
} 
\end{center} 
\end{table} 

Using {orbital overlaps} in Table I, it is easy to evaluate the many-body overlap and Hamiltonian 
matrix element,
\begin{eqnarray}
\langle\Psi_{{\delta}}|\Psi_{{\delta'}}\rangle
&=&
\left(\frac{2b_\perp b_\perp'}{b_\perp^2+b_\perp'^2}\right)^{{24}}
\left(\frac{2b_z b_z'}{b_z^2+b_z'^2}\right)^{{12}}, 
\label{eq:ov}
\\
\frac{\langle\Psi_{{\delta}}|H|\Psi_{{\delta'}}\rangle}
{\langle\Psi_{{\delta}}|\Psi_{{\delta'}}\rangle}
&=&\frac{{24}\hbar^2}{2m}\,
\left(\frac{2}{b_\perp^2+b_\perp'^2}
+\frac{1}{b_z^2+b_z'^2}\right) \nonumber \\
&&+\frac{1}{2}m\omega_0^2
\left(\frac{{48}b_\perp^2b_\perp'^2}{b_\perp^2+b_\perp'^2}
+\frac{{24}b_z^2b_z'^2}{b_z^2+b_z'^2}\right) \nonumber \\
&&+\frac{\kappa}{2}
\left[24
\left(\frac{b_\perp^2b_\perp'^2}{b_\perp^2+b_\perp'^2}
-\frac{b_z^2b_z'^2}{b_z^2+b_z'^2}\right)\right]^2. \nonumber \\
\label{eq:Hov}
\end{eqnarray}
The result for the overlap can also be obtained  as a special case of the
general formula for many-particle harmonic oscillator configurations 
given in Ref. \cite[Eq.3.5]{arima1959}. 
\footnote{The variable $\varepsilon$ in the notation of Ref.
\cite{arima1959} is identical $(\delta - \delta')/6$.}

Finally we expand Eq.  (\ref{eq:Hov}) to the quadratic order of $\delta$ and
$\delta'$ to obtain
\begin{equation}
\frac{\langle\Psi_{{\delta}}|H|\Psi_{{\delta'}}\rangle}
{\langle\Psi_{{\delta}}|\Psi_{{\delta'}}\rangle} =
18\hbar\omega_0-\hbar\omega_0(\delta-\delta')^2+ ... 
\label{eq:Hov2}
\end{equation}
One thus finds 
\begin{equation}
h_2=\hbar\omega_0.
\label{h2-def}
\end{equation}
\begin{table}[htb] 
\begin{center} 
\begin{tabular}{c|c} 
\hline
\hline 
$\psi_{000}$ & $-{8}\hbar\omega_0[2(a_2^{(\perp)})^2+(a_2^{(z)})^2]$ \\
$\psi_{001}$ & $-{8}\hbar\omega_0[2(a_2^{(\perp)})^2+(a_3^{(z)})^2]$ \\
$\psi_{100}$ & $-{8}\hbar\omega_0[(a_3^{(\perp)})^2+(a_2^{(\perp)})^2
+(a_2^{(z)})^2]$ \\
$\psi_{010}$ & $-{8}\hbar\omega_0[(a_3^{(\perp)})^2+(a_2^{(\perp)})^2
+(a_2^{(z)})^2]$ \\
\hline
\hline
\end{tabular} 
\caption{
The contribution of each single-particle wave function $\psi_k$ 
to the second order term 
in the energy overlap, $-\sum_{ph}|C_{ph}|^2(\epsilon_p-\epsilon_h)$. 
Here, the 
particle-hole energy is 
$\epsilon_p-\epsilon_h=2\hbar\omega_0$, 
and the spin degeneracy is also taken into account. 
}
\end{center} 
\end{table} 

We now evaluate $h_2$ in the orbital energy formula Eq. (\ref{orbital-eq2})
to compare with
Eq. (\ref{h2-def}).  We apply the formula to the matrix elements between
 $\Psi_{{\delta}}$ and $\Psi_{{-\delta}}$ taking 
$\Psi_{{0}}$ as the reference configuration.  The main task is to
determine the coefficients $c_{ph}$ of the deformed orbits in
terms of the orbitals in the spherical basis.  
Since we are interested in the second order behavior of the Hamiltonian 
overlap with respect to $\delta$, it is sufficient to expand the wave functions 
up to 
$n=2$ and $n=3$ for $\phi_0(b,x)$ and $\phi_1(b,x)$, respectively. 
This leads to 
\begin{eqnarray}
\phi_0(b,x)&\propto&\phi_0(b_0,x)+a_2\phi_2(b_0,x)+\cdots \\
\phi_1(b,x)&\propto&\phi_1(b_0,x)+a_3\phi_3(b_0,x)+\cdots 
\end{eqnarray}
with the coefficients of 
\begin{equation}
a_2=
\frac{1}{\sqrt{2}}\,\frac{b^2-b_0^2}{b^2+b_0^2}, 
\end{equation}
and
\begin{equation}
a_3=\sqrt{\frac{3}{2}}\,
\frac{b^2-b_0^2}{b^2+b_0^2}  
\end{equation}
for the one-dimensional Gaussians in an orbital.
The three-dimensional orbitals  $\phi_{n_x,n_y,n_z}$  are expanded as
\begin{eqnarray}
\phi_{000}&\propto& 
\phi_{000}^{(0)}+a_2^{(\perp)}(\phi_{200}^{(0)}+\phi_{020}^{(0)})
+a_2^{(z)}\phi_{002}^{(0)}+\cdots \nonumber \\
\\
\phi_{001}&\propto& 
\phi_{001}^{(0)}+a_2^{(\perp)}(\phi_{201}^{(0)}+\phi_{021}^{(0)})
+a_3^{(z)}\phi_{003}^{(0)}+\cdots \nonumber \\
\\
\phi_{100}&\propto& 
\phi_{100}^{(0)}+a_3^{(\perp)}\phi_{300}^{(0)}+
a_2^{(\perp)}\phi_{120}^{(0)}+
a_2^{(z)}\phi_{102}^{(0)}+\cdots \nonumber \\
\end{eqnarray}
where 
$\phi_k^{(0)}$ is a spherical orbital.
The contribution of each single-particle state 
to the 
second order term 
in the energy overlap, $-\sum_{ph}|C_{ph}|^2(\epsilon_p-\epsilon_h)$, 
is shown in Table 2. 
From this table, one has 
\begin{eqnarray}
&&-\sum_{ph}|C_{ph}|^2(\epsilon_p-\epsilon_h) \nonumber \\
&=&
-{8}\hbar\omega_0[6(a_2^{(\perp)})^2+2(a_3^{(\perp)})^2+3(a_2^{(z)})^2+(a_3^{(z)})^2] 
\nonumber \\
\\
&=&
-{8}\hbar\omega_0\left\{
6\left(\frac{b_\perp^2-b_0^2}{b_\perp^2+b_0^2}\right)^2
+3\left(\frac{b_z^2-b_0^2}{b_z^2+b_0^2}
\right)^2
\right\}. 
\end{eqnarray}
Expanding this quantity up to the second order of $\delta$, one finds 
\begin{equation}
-\sum_{ph}|C_{ph}|^2(\epsilon_p-\epsilon_h) \sim 
-{4}\hbar\omega_0\delta^2. 
\end{equation}
This coincides with the second order term in Eq. 
(\ref{eq:Hov2}) with $\delta'=-\delta$. 
It is easy to confirm that the same relation holds also with a deformed reference 
configuration, with $b_0=b(\delta_{\rm ref})$ and 
$b=b(\delta_{\rm ref}+\delta)$. 

\section{Summary}

We have derived a simple formula for the Hamiltonian matrix elements 
by the Generator Coordinate Method. 
The formula is based on a residual interaction of separable form, determined
by a diabatic treatment of the generator coordinate.  The formula was shown
to be exact for the leading dependence on $q_1-q_2$ for two solvable
models.

{Although it is approximate, the formula has several attractive features. First, it does not 
require full details of the many-body Hamiltonian. Thus 
one can carry out GCM calculations even when only the 
mean-field potential is known. This is the case when using a phenomenological 
mean-field potential such as a Woods-Saxon potential.
Second, the formula is much simpler than the original multi-step procedure
based on the 
generalized Wick theorem or the L\"owdin formula. 
Third, by using a separable interaction, one can avoid the well-known 
difficulties of treating energy functionals as Hamiltonians.
}

We plan to report numerical calculations of more realistic models in
a separate 
publication.

\section*{Acknowledgements}
We thank P.-G. Reinhard for comments on the manuscript.
This work was supported in part by
JSPS KAKENHI Grant Numbers JP19K03861 and JP21H00120.


\begin{thebibliography}{100}

\bibitem{RS}
P. Ring and P. Schuck,
{\it The Nuclear Many Body Problem}
(Springer-Verlag, New York, 1980).

\bibitem{RG87}
P.-G. Reinhard and K. Goeke, 
Rep. Prog. in Phys. {\bf 50}, 1 (1987). 

\bibitem{RRR18}
L.M. Robledo, T.R. Rodriguez, and R.R. Rodriguez-Guzman, 
J. of Phys. G{\bf 46}, 013001 (2019). 

\bibitem{BB} R. Balian and E. Br\'ezin, Nuovo Cimento {\bf 64B}, 34 (1969).

\bibitem{bo90} P.~Bonche et al., Nucl. Phys. {\bf A510}, 466 (1990). 

\bibitem{ba13} M~Baldo, L.M.~Robledo, P.~Schuck and X.
Vinas, Phys. Rev.
C {\bf 87}, 064305 (2013).

\bibitem{ro12} R.~Roth, S.~Binder, K.~Vobig, et al., Phys. Rev. Lett.
{\bf 109}, 052501 (2012).

\bibitem{thouless1960}
D.J. Thouless, Nucl. Phys. {\bf 21}, 225 (1960). 

\bibitem{bohr-mottelson}
A. Bohr and B.R. Mottelson, {\it Nuclear Structure}, Vol. II (Benjamin, New York, 1975). 

\bibitem{bertsch1983}
G.F. Bertsch, Prog. Theo. Phys. Suppl. {\bf 74\&75}, 115 (1983). 

\bibitem{su81} T.~Suzuki and H.~Sagawa, Prog. Theo. Phys. {\bf 65}, 565
(1981).

\bibitem{tagami2013}
S. Tagami, Y.R. Shimizu, and J. Dudek, 
Phys. Rev. C{\bf 87}, 054306 (2013). 

\bibitem{alhassid} Y.~Alhassid, L.~Fang, and H.~Nakada, Phys. Rev. Lett.
{\bf 101}, 082501 (2008).

\bibitem{lowdin1955}
{P.O. L\"owdin, Phys. Rev. {\bf 97}, 1490 (1955). }

\bibitem{brink1965}
{D.M. Brink, {\it Proc. International School of Physics Enrico Fermi'} course XXXVI  (1965), p. 247.}

\bibitem{fiolhais83}
{C. Fiolhais and R.M. Dreizler, Nucl. Phys. {\bf A393}, 205 (1983). }

\bibitem{be19}G.F.~Bertsch and W.~Younes, Ann. Phys. {\bf 403}, 68 (2019).

\bibitem{arima1959}
A. Arima and S. Yoshida, Nucl. Phys. {\bf 12}, 139 (1959). 

\end{thebibliography}
\end{document}